\documentclass{birkjour}

\usepackage[T1]{fontenc}
\usepackage[utf8]{inputenc}
\usepackage{lmodern}
\usepackage{microtype}
\usepackage{lipsum} 

\usepackage{mathtools}
\usepackage{amsthm}
\usepackage{amssymb}
\usepackage{bm}

\usepackage{booktabs}
\usepackage{array}
\usepackage{multirow}
\usepackage{makecell}
\usepackage{float}

\usepackage{graphicx}
\usepackage{xcolor}
\usepackage{tikz}
\usepackage{colortbl}
\usepackage{subcaption}
\usetikzlibrary{decorations.pathreplacing, calc, intersections}
\usetikzlibrary{arrows.meta}

\usepackage{listings}
\usepackage[colorlinks=true,
            linkcolor=blue!60!black,
            citecolor=blue!60!black,
            urlcolor=blue!60!black]{hyperref}
\usepackage[capitalise, noabbrev]{cleveref}

\newtheorem{theorem}{Theorem}[section]

\theoremstyle{definition}
\newtheorem{definition}[theorem]{Definition}

\theoremstyle{remark}
\newtheorem{remark}[theorem]{Remark}

\definecolor{C0}{RGB}{195, 10,  58} 
\definecolor{C1}{RGB}{ 20,130, 200} 
\definecolor{C2}{RGB}{  0,168, 110} 
\definecolor{C3}{RGB}{210,160,   0} 

\begin{document}

\title{Generalize cross-ratios in n-dimensional Plane-Based Geometric Algebra}

\author[E. Harquin \and S. Breuils \and P. Monasse \and V. Biri \and V. Nozick]{Enzo Harquin \and St\'ephane Breuils \and Pascal Monasse \and Venceslas Biri \and Vincent Nozick}

\address{%
  Enzo Harquin, Pascal Monasse, Venceslas Biri, Vincent Nozick\\
  Universit\'e Gustave Eiffel, LIGM,\\
  Champs-sur-Marne, F-77420, France\\
  \texttt{\{enzo.harquin2,venceslas.biri,pascal.monasse,vincent.nozick\}@univ-eiffel.fr}
  \and\\
  St\'ephane Breuils\\
  Universit\'e Savoie Mont-Blanc, CNRS, LAMA,\\
  Chamb\'ery, F-73000, France\\
  \texttt{stephane.breuils@univ-smb.fr}
}



\begin{abstract}
We develop a complete theory of projective cross-ratios in
$n$-dimensional Plane-Based Geometric Algebra (PGA),
$\mathbb{R}_{n,0,1}$, covering geometric objects of every grade:
finite and ideal points, hyperplanes, and intermediate flats.
For each object type and configuration, we establish an explicit
cross-ratio formula, prove that it recovers the appropriate
classical invariant, and identify the canonical pairwise
measurement operator.
A systematic duality analysis further revealed that all eight
configurations organize into four dual pairs under the Hodge dual,
and that all measurement operators reduce to either the commutator
or the commutator dual, depending solely on
the geometric configuration rather than on object grade.
In each case the formula recovers the appropriate classical invariant:
signed distance ratios for parallel configurations and sine
cross-ratios for secant ones.
These results establish the cross-ratio as a grade-agnostic projective invariant within PGA, and provide a constructive foundation for defining $n$-dimensional homographies directly from prescribed invariants.
\end{abstract}

\maketitle


\section{Introduction}
\label{sec:intro}
The cross-ratio is the fundamental invariant of projective geometry:
any scalar quantity associated with four collinear points that is
preserved by every projective transformation can be expressed as a
function of the cross-ratio~\cite{stillwell}. Its importance extends
well beyond classical projective geometry: it governs homographic
relations in computer vision~\cite{hartley2003multiple}, underpins
reconstruction algorithms in multi-view geometry, and appears
naturally in any setting where projective transformations act on
geometric objects.

Plane-Based Geometric Algebra
(PGA)~\cite{gunn2011geometrykinematicsrigidbody,
gunn2011homogeneousmodeleuclideangeometry} offers a compelling
algebraic framework for projective geometry: hyperplanes are the
primitive elements, all geometric objects arise as blades of specific
grades, and Euclidean isometries admit compact versor representations
as motors. However, the cross-ratio has received limited attention in
this setting. The present paper fills this gap by developing a
complete, grade-agnostic theory of projective cross-ratios within
$n$-dimensional PGA $\mathbb{R}_{n,0,1}$.

\subsection{Motivation and Related Work}
\label{sec:motivation}
The cross-ratio has been studied in the context of geometric algebra
in~\cite{bayro2002geometric,lasenby1999analysis} and, in the form of lecture notes, in~\cite{lundholm2009clifford}
where formulas are derived for collinear points and applied to
automatic object recognition in real-world scenes. Those treatments,
however, are restricted to point configurations and formulated in
non-degenerate algebras, which do not
distinguish finite from ideal elements and offer no natural
representation of parallelism or elements at infinity. As a
consequence, the connection between the point cross-ratio and the
angle cross-ratio for pencils of lines, the behavior of the invariant
under duality, and its extension to intermediate grades such as lines
or planes have not been addressed in that framework.

PGA, by contrast, is specifically designed to handle the
finite--ideal distinction through its degenerate metric, and its
grade structure naturally encodes flat objects of every dimension.
This makes it the right setting for a unified treatment. The present
work derives cross-ratio formulas for all object types in
$n$-dimensional PGA and shows that all formulas reduce to a small
family of operators determined solely by the geometric configuration
of the four objects.

\subsection{Contributions}
\label{sec:contributions}

The main contributions of this paper are:
\begin{itemize}
  \item Cross-ratio formulas for geometric objects of every grade in
    $n$-dimensional PGA, covering finite and ideal points,
    hyperplanes, and intermediate flats, with explicit treatment of
    parallel and secant configurations (\cref{sec:pga-cr}).
  \item A systematic duality analysis showing that all eight
    configurations organize into four dual pairs under the Hodge
    dual, and that the canonical pairwise measurement operators
    reduce uniformly to the commutator or its dual \cref{sec:recap}).
\end{itemize}

\subsection{Notation and Conventions}
\label{sec:notation}

We work throughout in $n$-dimensional PGA $\mathbb{R}_{n,0,1}$,
whose construction and products are recalled in
\cref{sec:background}. Arbitrary multivectors are denoted by
italic capital letters $A$ and italic lowercase letters $a$ for $1$-vector. Geometric objects are denoted by bold
symbols: $\mathbf{P}$ for finite points, $\mathbf{V}$
for ideal points, $\boldsymbol{\Pi}$ for hyperplanes, and $\mathbf{F}$ for intermediate flats (i.e. lines in 3D PGA).
The Hodge dual of a multivector $A$ is written
$A^\star$. The cross-ratio associated to four objects $A_1, A_2, A_3, A_4$ is written $\{A_1, A_2; A_3, A_4\}$.

\section{Background: Plane-Based Geometric Algebra}
\label{sec:background}

Plane-Based Geometric Algebra (PGA), 
developed by Gunn~\cite{gunn2011geometrykinematicsrigidbody,
gunn2011homogeneousmodeleuclideangeometry}, is a Clifford algebra designed
for the algebraic representation of projective geometry, in which hyperplanes are the primitive elements and points arise as their intersections. This section introduces $n$-dimensional PGA to the extent needed to follow the contributions of this paper. For a more thorough treatment, we refer the reader to~\cite{dorst2022guidedPGA,
dorstmaytheforquebewithyou,leger2025tutorial}.

\subsection{The Algebra \texorpdfstring{$\mathbb{R}_{n,0,1}$}{G(n,0,1)}: Metric and Pseudoscalar}
\label{sec:algebra}

The $n$-dimensional PGA is the Clifford algebra
$\mathbb{R}_{n,0,1}$ generated by $n+1$ basis vectors
$\mathbf{e}_0, \mathbf{e}_1, \ldots, \mathbf{e}_n$ subject to the
degenerate symmetric bilinear form
\begin{equation}
\begin{aligned}
  \mathbf{e}_i \cdot \mathbf{e}_i &= 1~\text{for}~i = 1, 2, \ldots, n \\
  \mathbf{e}_0 \cdot \mathbf{e}_0 &= 0 \\
  \mathbf{e}_i \cdot \mathbf{e}_j &= 0~\text{for}~i \neq j
\end{aligned}
\end{equation}
The generator $\mathbf{e}_0$ is therefore a null vector $\mathbf{e}_0^2 = 0$, it encodes the \emph{ideal hyperplane} at infinity. The unit pseudoscalar of the full algebra and its Euclidean counterpart are
\begin{equation}
  \mathbf{I} = \mathbf{e}_{012\cdots n},
  \qquad
  \mathbf{I}_E = \mathbf{e}_{12\cdots n}, 
  \qquad
  \mathbf{I} = \mathbf{e}_0 \mathbf{I}_E.
\end{equation}
The origin point of $n$-dimensional PGA corresponds to the Euclidean pseudoscalar, denoted $\mathbf{O} = \mathbf{I}_E$.

\subsection{The Geometric Product and Commutator}
\label{sec:products}

The fundamental operation of Geometric Algebra is the \emph{geometric product}. For any two \emph{1-vectors} $a, b$, it decomposes exactly into a commutative and an anti-commutative part:
\begin{equation}
  ab = a \cdot b + a \wedge b
\end{equation}
For general multivectors of arbitrary grade, this clean two-term decomposition no longer holds. In the case where $A$ is of grade $r$ and $B$ of grade $s$ blades, the geometric product produces a sum of blades of grades $|r-s|, |r-s|+2, \ldots, r+s$, restricted to the maximum grade $n+1$. Again, the decomposition splits into a commutative part and an anti-commutative part:
\begin{equation}
\label{eq:gp}
  AB = \tfrac{1}{2}(AB + BA)
     + \tfrac{1}{2}(AB - BA)
\end{equation}
Specifically, a grade $k$ belongs to the commutative part $\tfrac{1}{2}(AB+BA)$ if and only if
\begin{equation}
\label{eq:sym_condition}
  \frac{k(k-1)}{2} + \frac{r(r-1)}{2} + \frac{s(s-1)}{2} \equiv 0 \pmod{2},
\end{equation}
and to the anti-commutative part $\tfrac{1}{2}(AB-BA)$ otherwise.

~\\The \emph{commutator product} is defined as the anti-commutative part,
\begin{equation}
  A \times B = \tfrac{1}{2}(AB - BA).
\end{equation}
The commutator selects precisely the grades for which condition~\eqref{eq:sym_condition}
fails, up to the maximum grade $n+1$ of the algebra. For better numerical performances, it is more efficient to compute $AB$ once and extract the desired grade components.
It follows that for any two \emph{1-vectors} $a, b$, the commutator coincides with the wedge
\begin{equation}
  a \times b = a \wedge b
\end{equation}

\subsection{Finite and Ideal Objects: The Euclidean Split}
\label{sec:split}

Any multivector $A$ decomposes uniquely as
\begin{equation}
  A = A_E + \mathbf{e}_0 A_I,
\end{equation}
where $A_E$ is the \emph{Euclidean part}, containing no factor $\mathbf{e}_0$,
and $A_I$ is the \emph{ideal part}. For instance, in $3$-dimensional PGA:
\begin{align}
    \nonumber
    & \underbrace{a_1\mathbf{e}_1 + a_2\mathbf{e}_2 + a_3\mathbf{e}_3}_{\text{Euclidean}}
     + \mathbf{e}_0 \underbrace{a_0}_{\text{Ideal}}, \\[4pt]
    \nonumber
    & \underbrace{a_0\mathbf{e}_{123}}_{\text{Euclidean}}
     + \mathbf{e}_0 \underbrace{%
         (-a_1\mathbf{e}_{23} + a_2\mathbf{e}_{13} - a_3\mathbf{e}_{12})%
       }_{\text{Ideal}}.
\end{align}
The Euclidean part encodes the object's orientation or weight, while
the ideal part encodes its position relative to the origin.

An object with a nonzero Euclidean part is called \emph{finite}; an object with a vanishing Euclidean part is called \emph{ideal}. An object with a vanishing ideal part necessarily passes through the origin.
For further details, refer to \cite{dorstmaytheforquebewithyou}.

\subsection{The Hodge Dual and Dual Operators}
\label{sec:dual}

\paragraph{Hodge Dual}
In PGA, the dual is defined via the Hodge dual, see \cite{gunn2011geometrykinematicsrigidbody}.
The Hodge dual maps each basis blade $\mathbf{e}_J$ to the unique
blade $\mathbf{e}_J^\star$ such that, by convention
\begin{equation}
  \mathbf{e}_J \wedge \mathbf{e}_J^\star = \mathbf{I},
\end{equation}
and is extended to all multivectors by linearity.
Using the Euclidean split, it admits the closed-form expression
\begin{equation}
  A^\star = \widetilde{A_I}\,\mathbf{I}_E
          + \mathbf{e}_0\,\widehat{\widetilde{A_E}}\,\mathbf{I}_E,
  \label{eq:dual}
\end{equation}
where $\widetilde{\cdot}$ and $\widehat{\cdot}$ denote respectively the reverse and the grade involution,
two canonical anti-automorphisms which act on multivectors, see \cite{LeoBook}.

\paragraph{Dual Operators}
Dual operators are typically introduced in the point-based literature. We include them here explicitly, as they provide access to a wider family of algebraic operations central to the present work.

\noindent
The \emph{regressive product}
(also called \emph{join}) is the dualized wedge product:
\begin{equation}
  A \vee B = \bigl(A^\star \wedge B^\star\bigr)^\star.
\end{equation}
The \emph{inner dual} is the dualized inner product:
\begin{equation}
  A \cdot^\star B = \bigl(A^\star \cdot B^\star\bigr)^\star.
\end{equation}
The \emph{commutator dual} is the dualized commutator product:
\begin{equation}
  A \times^\star B = \bigl(A^\star \times B^\star\bigr)^\star.
\end{equation}

It follows that for any two \emph{n-vectors} $A, B$, the commutator dual coincides with the regressive product
\begin{equation}
  A \vee B = A \times^\star B.
\end{equation}

\subsection{Geometric Objects: Hyperplanes, Flats, and Points}
\label{sec:objects}

Geometric objects in $n$-dimensional PGA are represented by blades of
specific homogeneous grades.

\begin{itemize}
  \item \textbf{Grade $1$ - Hyperplanes.}
    A hyperplane is defined as a $1$-vector.
    \begin{equation}
    \boldsymbol{\Pi} = \underbrace{a_1\mathbf{e}_1 + a_2\mathbf{e}_2 + \cdots + a_n\mathbf{e}_n}_{\text{encodes normal}}
     + \underbrace{a_0 \mathbf{e}_0}_{\text{encodes distance}}.
   \end{equation}
   $\mathbf{e}_0$ encodes the unique ideal hyperplane. 

  \item \textbf{Grade $k$ ($1 < k < n$) - Intermediate flats.}
    Flats are lines in $3$d, planes in $4$d, and their higher-dimensional analogues. Below an example of lines in $3$-dimensional PGA.
    \begin{equation}
    (3\text{d})~ \mathbf{F} = \underbrace{a_{12}\mathbf{e}_{12} + a_{13}\mathbf{e}_{13}+ a_{23}\mathbf{e}_{23}}_{\text{encodes direction}}
     + \underbrace{a_{01} \mathbf{e}_{01} + a_{02} \mathbf{e}_{02} + a_{03} \mathbf{e}_{03}}_{\text{encodes moment}}.
   \end{equation}
   
  \item \textbf{Grade $n$ - Points.}
    A point $\mathbf{P}$ is defined as an $n$-vector, the dual of a $1$-vector.
    \begin{equation}
    \mathbf{P} = \big(\underbrace{a_1\mathbf{e}_1 + a_2\mathbf{e}_2 + \cdots + a_n\mathbf{e}_n}_{\text{encodes position}}
     + \underbrace{a_0 \mathbf{e}_0}_{\text{encodes weight}}\big)^\star.
   \end{equation}
    An ideal point corresponds to $a_0 = 0$. The dual $\mathbf{e}_0^\star = \mathbf{I}_E$ encodes the unique origin.
\end{itemize}
Note that points and hyperplanes are linked by duality, $\boldsymbol{\Pi} = \mathbf{P}^\star$. More generally, objects of grade-$k$ are linked to their dual of grade-$(n+1-k)$, where $n$ is the dimension of the Euclidean space.

\section{The Classical Cross-Ratio}
\label{sec:classical}

The cross-ratio is the fundamental projective invariant of the
projective line~$\mathbb{P}^1$: any scalar quantity associated with
four collinear points (or four concurrent lines) that is preserved by
every projective transformation can be expressed as a function of the
cross-ratio, see \cite[Sect.~5.8]{stillwell}. In this sense it generates all projective invariants of four points on a line, see \cite[Sect.~2.5]{hartley2003multiple}.
In this section we recall two classical forms of the cross-ratio.

\subsection{Cross-Ratio on the Projective Line}
\label{sec:proj-cr}

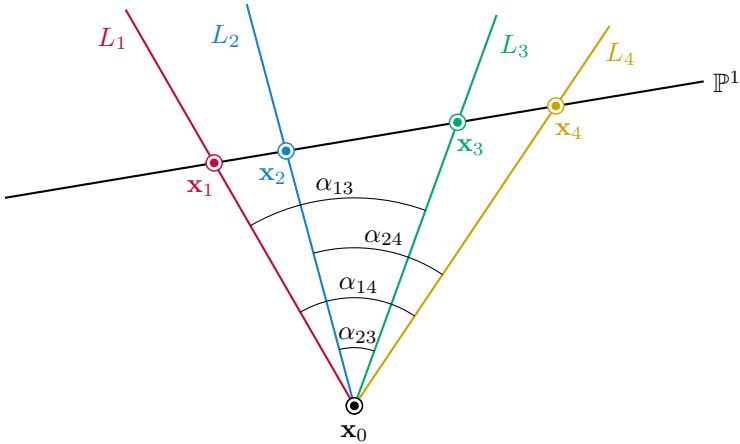
\begin{figure}[ht]
\centering
\begin{tikzpicture}[scale=1.1]
    \coordinate (O) at (0.0, -1.5);
    \fill[black] (O) circle (3pt);
    \draw[thick, name path=transversal] (-4.2, 1) -- (4.2, 2.4);
    \node[right, font=\normalsize] at (4.2, 2.4) {$\mathbb{P}^1$};
    \draw[C0, thick, name path=L1] (O) -- ++(120:5.5cm);
    \draw[C1, thick, name path=L2] (O) -- ++(105:5.0cm);
    \draw[C2, thick, name path=L3] (O) -- ++(70:5.0cm);
    \draw[C3, thick, name path=L4] (O) -- ++(56.1:5.5cm);
    \node[C0, font=\normalsize, left=2pt]  at ($(O)+(120:5.1cm)$)  {$L_1$};
    \node[C1, font=\normalsize, left=2pt]  at ($(O)+(105:4.6cm)$)  {$L_2$};
    \node[C2, font=\normalsize, right=2pt] at ($(O)+(70:4.6cm)$)   {$L_3$};
    \node[C3, font=\normalsize, right=2pt] at ($(O)+(56.1:5.1cm)$) {$L_4$};
    \path[name intersections={of=L1 and transversal, by=P1}];
    \path[name intersections={of=L2 and transversal, by=P2}];
    \path[name intersections={of=L3 and transversal, by=P3}];
    \path[name intersections={of=L4 and transversal, by=P4}];
    \fill[C0] (P1) circle (3pt);
    \fill[C1] (P2) circle (3pt);
    \fill[C2] (P3) circle (3pt);
    \fill[C3] (P4) circle (3pt);
    \fill[white] (P1) circle (2.5pt);
    \fill[white] (P2) circle (2.5pt);
    \fill[white] (P3) circle (2.5pt);
    \fill[white] (P4) circle (2.5pt);
    \fill[C0] (P1) circle (1.5pt);
    \fill[C1] (P2) circle (1.5pt);
    \fill[C2] (P3) circle (1.5pt);
    \fill[C3] (P4) circle (1.5pt);
    \fill[black] (O) circle (3pt);
    \fill[white] (O) circle (2.5pt);
    \fill[black] (O) circle (1.5pt);
    \node[C0, left=5pt,  below=3pt, font=\normalsize] at (P1) {$\mathbf{x}_1$};
    \node[C1, left=5pt,  below=3pt, font=\normalsize] at (P2) {$\mathbf{x}_2$};
    \node[C2, right=5pt, below=3pt, font=\normalsize] at (P3) {$\mathbf{x}_3$};
    \node[C3, right=5pt, below=3pt, font=\normalsize] at (P4) {$\mathbf{x}_4$};
    \node[below=4pt, font=\normalsize] at (O) {$\mathbf{x}_0$};
    \draw[black, thin] (O) ++(105:0.7cm) arc (105:70:0.7cm);
    \node[black, font=\normalsize] at ($(O)+(87:0.85cm)$) {$\alpha_{23}$};
    \draw[black, thin] (O) ++(120:1.3cm) arc (120:56.1:1.3cm);
    \node[black, font=\normalsize] at ($(O)+(88:1.45cm)$) {$\alpha_{14}$};
    \draw[black, thin] (O) ++(105:1.9cm) arc (105:56.1:1.9cm);
    \node[black, font=\normalsize] at ($(O)+(80:2.05cm)$) {$\alpha_{24}$};
    \draw[black, thin] (O) ++(120:2.5cm) arc (120:70:2.5cm);
    \node[black, font=\normalsize] at ($(O)+(95:2.65cm)$) {$\alpha_{13}$};
\end{tikzpicture}
\caption{The classical cross-ratio in its two equivalent forms. Four
lines $L_1, L_2, L_3, L_4$ meet at $\mathbf{x}_0$ and intersect the
transversal $\mathbb{P}^1$ in four points $\mathbf{x}_1, \mathbf{x}_2,
\mathbf{x}_3, \mathbf{x}_4$. The cross-ratio equals both the ratio of
signed distances along $\mathbb{P}^1$ and the ratio of sines of the
angles $\alpha_1, \alpha_2, \alpha_3, \alpha_4$ at $\mathbf{x}_0$.}
\label{fig:crossratio-unified}
\end{figure}

Given four distinct points in homogeneous coordinates $\mathbf{x}_1, \mathbf{x}_2, \mathbf{x}_3, \mathbf{x}_4 \in \mathbb{P}^1$, with $\mathbf{x}_i = (x_i, w_i)^\top$, their cross-ratio is defined as
\begin{equation}
  \{\mathbf{x}_1, \mathbf{x}_2; \mathbf{x}_3, \mathbf{x}_4\}
  = \frac{|\mathbf{x}_1\mathbf{x}_3|\,|\mathbf{x}_2\mathbf{x}_4|}
         {|\mathbf{x}_1\mathbf{x}_4|\,|\mathbf{x}_2\mathbf{x}_3|},
\end{equation}
where $|\mathbf{x}_i\mathbf{x}_j|$ denotes the $2\times 2$ determinant
\begin{equation}
  |\mathbf{x}_i\mathbf{x}_j|
  = \det\begin{bmatrix} x_i & x_j \\ w_i & w_j \end{bmatrix}
  = x_i w_j - x_j w_i.
\end{equation}
A visual example is illustrated in \cref{fig:crossratio-unified}.

The resulting invariant is independent of the choice of homogeneous representative. When all $w_i \neq 0$, writing $\hat{x}_i = x_i / w_i$, $\hat{\mathbf{x}}_i = (\hat{x}_i, 1)^\top$ and factoring $|\mathbf{x}_i \mathbf{x}_j| = w_i w_j (\hat{x}_i - \hat{x}_j)$, the common factor $w_1 w_2 w_3 w_4$ cancels and the cross-ratio reduces to
\begin{equation}
\label{eq:classical-cr}
  \{\mathbf{x}_1, \mathbf{x}_2; \mathbf{x}_3, \mathbf{x}_4\}
  = \frac{(\hat{x}_1 - \hat{x}_3)(\hat{x}_2 - \hat{x}_4)}
         {(\hat{x}_1 - \hat{x}_4)(\hat{x}_2 - \hat{x}_3)}.
\end{equation}
When one of the four points, say $\mathbf{x}_4$, is sent to infinity
along the line, the cross-ratio degenerates into a simple ratio,
\begin{equation}
  \{\hat{\mathbf{x}}_1, \hat{\mathbf{x}}_2; \hat{\mathbf{x}}_3, \mathbf{x}_4\}
  \xrightarrow{\mathbf{x}_4 \to \infty}
  \frac{|\hat{\mathbf{x}}_1\hat{\mathbf{x}}_3|}{|\hat{\mathbf{x}}_2\hat{\mathbf{x}}_3|},
\end{equation}
which is the \emph{affine ratio} of the triple
$(\hat{\mathbf{x}}_1, \hat{\mathbf{x}}_2, \hat{\mathbf{x}}_3)$. This quantity is invariant under
affine transformations but not under the full projective group.
Unlike the full cross-ratio, the affine ratio is not independent of the choice of homogeneous representative; the three points must be taken in normalized form $(w_i=1)$.

\subsection{The Angle Cross-Ratio for Pencils of Lines}
\label{sec:sine-cr}

When the four points are replaced by four concurrent lines, the
cross-ratio takes an angular form. Let $L_1, L_2, L_3, L_4$ be four lines in the plane, meeting at a common point $\mathbf{x}_0$, and let
$\mathbf{x}_1, \mathbf{x}_2, \mathbf{x}_3, \mathbf{x}_4 \in \mathbb{P}^2$ be any four homogeneous points lying respectively on these lines, as depicted in \cref{fig:crossratio-unified}. Their cross-ratio is
\begin{equation}
  \{L_1, L_2; L_3, L_4\}
  = \frac{|\mathbf{x}_0\mathbf{x}_1\mathbf{x}_3|\,|\mathbf{x}_0\mathbf{x}_2\mathbf{x}_4|}
         {|\mathbf{x}_0\mathbf{x}_1\mathbf{x}_4|\,|\mathbf{x}_0\mathbf{x}_2\mathbf{x}_3|},
\end{equation}
where $|\mathbf{x}_0\mathbf{x}_i\mathbf{x}_j|$ denotes the $3\times 3$
determinant whose columns are the homogeneous coordinates of
$\mathbf{x}_0$, $\mathbf{x}_i$, and $\mathbf{x}_j$. This determinant is proportional to the signed area of the triangle $(\mathbf{x}_0, \mathbf{x}_i, \mathbf{x}_j)$, which
factors as
\begin{equation}
  |\mathbf{x}_0\mathbf{x}_i\mathbf{x}_j|
  = \|\mathbf{x}_0\mathbf{x}_i\|\,\|\mathbf{x}_0\mathbf{x}_j\|\,\sin(L_i, L_j),
\end{equation}
where $\sin(L_i, L_j)$ denotes the sine of the angle
between lines $L_i$ and $L_j$. Substituting into the cross-ratio, all
distance factors cancel, yielding the \emph{sine cross-ratio}:
\begin{equation}
  \{L_1, L_2; L_3, L_4\}
  = \frac{\sin(L_1, L_3)\,\sin(L_2, L_4)}
         {\sin(L_1, L_4)\,\sin(L_2, L_3)}
  = \frac{\sin(\alpha_{13}) \, \sin(\alpha_{24})}{\sin(\alpha_{14}) \, \sin(\alpha_{23})},
\end{equation}
where $\alpha_{ij}$ denotes the angle between $L_i$ and $L_j$.

The sine cross-ratio coincides with the point cross-ratio; this equality is expressed by the identity
\begin{equation}
  \{L_1, L_2; L_3, L_4\}
  = \{\mathbf{x}_1, \mathbf{x}_2; \mathbf{x}_3, \mathbf{x}_4\},
\end{equation}
where $\mathbf{x}_1, \mathbf{x}_2, \mathbf{x}_3, \mathbf{x}_4$ are the four intersection points of the lines with any transversal line in $\mathbb{P}^2$.

\begin{remark}
Although defined in $\mathbb{P}^2$, the angle cross-ratio is intrinsically a $\mathbb{P}^1$ invariant, since lines through a common point form a pencil isomorphic to $\mathbb{P}^1$, as illustrated in \cref{fig:crossratio-unified}.
Thus, the invariant depends only on this one-parameter family, not on the ambient projective space.
\end{remark}


\section{Cross-Ratio in PGA}
\label{sec:pga-cr}
This section develops cross-ratio formulas for all geometric object
types in $n$-dimensional PGA. We proceed systematically, starting
from finite points and extending to hyperplanes, ideal points, and
intermediate flats, each case arising naturally from the grade
structure and the degenerate metric of PGA. In each case, the
cross-ratio is defined as a ratio of pairwise measurements between
the four objects, and we show that the resulting invariant reduces to
the classical formulas of \cref{sec:proj-cr,sec:sine-cr}.

\subsection{Classical Point Cross-Ratio in PGA}
\label{sec:classicalcrossratio}
\begin{definition}
\label{def:cr-points-pga}
Let $\mathbf{P}_1, \mathbf{P}_2, \mathbf{P}_3, \mathbf{P}_4$ be four distinct collinear finite points in $n$-dimensional PGA. Their \emph{cross-ratio} is
\begin{equation}
  \{\mathbf{P}_1, \mathbf{P}_2; \mathbf{P}_3, \mathbf{P}_4\}
  \;=\;
  \frac{(\mathbf{P}_1 \vee \mathbf{P}_3)(\mathbf{P}_2 \vee \mathbf{P}_4)}
       {(\mathbf{P}_1 \vee \mathbf{P}_4)(\mathbf{P}_2 \vee \mathbf{P}_3)}.
  \label{eq:cr-points-pga}
\end{equation}
\end{definition}
Each regressive product $\mathbf{P}_i \vee \mathbf{P}_j$ is a grade-$(n-1)$ blade proportional to the unit support line $\hat{\mathbf{L}}$, with scalar factor equal to the signed parameter difference $t_i - t_j$ along the line:
\begin{equation}
  \mathbf{P}_i \vee \mathbf{P}_j
  = \|\mathbf{P}_i\|\,\|\mathbf{P}_j\|\,(t_i - t_j)\,\hat{\mathbf{L}}.
\end{equation}
Since all four pairs in~\eqref{eq:cr-points-pga} share the same
support line $\hat{\mathbf{L}}$, the geometric product
$\hat{\mathbf{L}}\hat{\mathbf{L}} = 1$ cancels for both numerator and
denominator, and the norms of the points cancel in the ratio.
Definition~\eqref{eq:cr-points-pga} therefore reduces to the
classical cross-ratio~\eqref{eq:classical-cr}:
\begin{equation}
  \{\mathbf{P}_1, \mathbf{P}_2; \mathbf{P}_3, \mathbf{P}_4\}
  = \frac{(t_1 - t_3)(t_2 - t_4)}{(t_1 - t_4)(t_2 - t_3)}.
\end{equation}
\begin{remark}
The formula~\eqref{eq:cr-points-pga} is consistent with the cross-ratio
expressions derived in~\cite{lasenby1999analysis,bayro2002geometric},
where the wedge product of grade-$1$ points is used in place of the regressive product of grade-$n$ points, the two being equivalent up to the choice of algebra.
\end{remark}

When one of the four points, say $\mathbf{P}_4$, is sent to infinity
along the line, it becomes an ideal point $\mathbf{V}_4$. The two
regressive products involving $\mathbf{V}_4$ then satisfy
\begin{equation}
  \mathbf{P}_i \vee \mathbf{V}_j
  = \|\mathbf{P}_i\|\|\mathbf{V}_j\|\,\hat{\mathbf{L}},
\end{equation}
so these factors cancel in the ratio, and~\eqref{eq:cr-points-pga}
degenerates to the affine ratio of the remaining triple:
\begin{equation}
  \label{eq:pga_affine_ratio}
  \{\mathbf{P}_1, \mathbf{P}_2; \mathbf{P}_3, \mathbf{V}_4\}
  = \frac{t_1 - t_3}{t_2 - t_3}.
\end{equation}
Again, this quantity is invariant under affine but not projective
transformations, consistently with the classical theory of
Section~\ref{sec:proj-cr}.

\begin{remark}
In Definition~\eqref{eq:cr-points-pga}, the geometric product in the numerator and denominator can be replaced by the inner product
\begin{equation}
  \{\mathbf{P}_1, \mathbf{P}_2; \mathbf{P}_3, \mathbf{P}_4\}
  =  \frac{(\mathbf{P}_1 \vee \mathbf{P}_3)\cdot(\mathbf{P}_2 \vee \mathbf{P}_4)}
       {(\mathbf{P}_1 \vee \mathbf{P}_4)\cdot(\mathbf{P}_2 \vee \mathbf{P}_3)},
\end{equation}
since all four factors $\mathbf{P}_i \vee \mathbf{P}_j$ are proportional to the same blade~$\hat{\mathbf{L}}$, so their geometric product reduces to a scalar, equal to their inner product.
\end{remark}

\subsection{Finite Points Dual: Hyperplanes Cross-Ratio}
\label{sec:dual-hyperplane}
The Hodge dual maps each point $\mathbf{P}_i$ to a hyperplane
$\boldsymbol{\Pi}_i = \mathbf{P}_i^\star$. Applied to the four
collinear points of \cref{sec:classicalcrossratio}, this duality
produces four hyperplanes whose common intersection is $\mathbf{L}^\star$, the dual of the common support line of the four points. Two cases arise depending on whether
$\mathbf{L}$ passes through the origin:
\begin{itemize}
  \item \textbf{Ideal intersection.}
    If $\mathbf{L}$ passes through the origin, $\mathbf{L}^\star$ is an ideal
    grade-$2$ flat, and the four dual hyperplanes share this ideal
    common intersection. They are therefore parallel.
  \item \textbf{Finite intersection.}
    If $\mathbf{L}$ does not pass through the origin, $\mathbf{L}^\star$ is a finite
    grade-$2$ flat, and the four dual hyperplanes meet along this
    common finite flat.
\end{itemize}

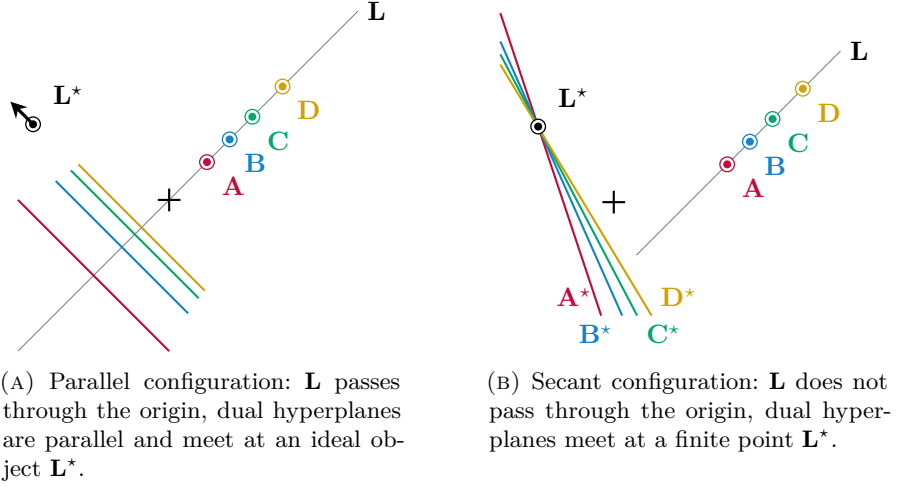
\begin{figure}[ht]
\centering
 
\begin{subfigure}[t]{0.45\textwidth}
\centering
\begin{tikzpicture}[scale=1]
    
    \coordinate (O) at (0, 0);
    
    \draw[gray, name path=lineL] (-2.0, -2.0) -- (2.5, 2.5);
    \node[right, font=\normalsize, black] at (2.5, 2.5) {$\mathbf{L}$};
    
    
    \draw[C0, thick] (-2.0, 0.0) -- (0.0, -2.0);    
    \draw[C1, thick] (-1.5, 0.25) -- (0.25, -1.5);  
    \draw[C2, thick] (-1.3, 0.39) -- (0.39, -1.3);  
    \draw[C3, thick] (-1.2, 0.47) -- (0.47, -1.2);  

    \coordinate (I) at (-1.8, 1);
    
    \coordinate (A) at (0.5, 0.5);
    \coordinate (B) at (0.8, 0.8);
    \coordinate (C) at (1.1, 1.1);
    \coordinate (D) at (1.5, 1.5);
    
    \fill[C0] (A) circle (3pt);
    \fill[white] (A) circle (2.5pt);
    \fill[C0] (A) circle (1.5pt);
    
    \fill[C1] (B) circle (3pt);
    \fill[white] (B) circle (2.5pt);
    \fill[C1] (B) circle (1.5pt);
    
    \fill[C2] (C) circle (3pt);
    \fill[white] (C) circle (2.5pt);
    \fill[C2] (C) circle (1.5pt);
    
    \fill[C3] (D) circle (3pt);
    \fill[white] (D) circle (2.5pt);
    \fill[C3] (D) circle (1.5pt);

    \fill[black] (I) circle (3pt);
    \fill[white] (I) circle (2.5pt);
    \fill[black] (I) circle (1.5pt);
    \draw[->, line width=1.2pt, -{Stealth[length=6pt, width=6pt]}] (I) -- ++(-0.3,0.3);

    \node[above right=4pt, font=\normalsize] at (I) {$\mathbf{L}^\star$};

    \node[C0, below right=2pt] at (A) {$\mathbf{A}$};
    \node[C1, below right=2pt] at (B) {$\mathbf{B}$};
    \node[C2, below right=2pt] at (C) {$\mathbf{C}$};
    \node[C3, below right=2pt] at (D) {$\mathbf{D}$};
    
    \draw[black, thick] (-0.15,0) -- (0.15,0);
    \draw[black, thick] (0,-0.15) -- (0,0.15);
    
\end{tikzpicture}
\caption{Parallel configuration: $\mathbf{L}$ passes through the origin, dual hyperplanes are parallel and meet at an ideal object~$\mathbf{L}^\star$.}
\label{fig:pga-duality-parallel}
\end{subfigure}
\hfill
\begin{subfigure}[t]{0.45\textwidth}
\centering
\begin{tikzpicture}[scale=1]
    
    \coordinate (O) at (0, 0);
    \draw[black, thick] (-0.15,0) -- (0.15,0);
    \draw[black, thick] (0,-0.15) -- (0,0.15);
    
    \draw[gray, name path=lineL] (0.3, -0.7) -- (3.0, 2.0);
    \node[right, font=\normalsize, black] at (3.0, 2.0) {$\mathbf{L}$};
    

    \def\ycut{-1.5}
    \draw[C0, thick] (-1.5, 2.5)   -- ({(\ycut+2)/-3}, \ycut);
    \draw[C1, thick] (-1.5, 2.125) -- ({(\ycut+1.25)/-2.25}, \ycut);
    \draw[C2, thick] (-1.5, 1.955) -- ({(\ycut+0.91)/-1.91}, \ycut);
    \draw[C3, thick] (-1.5, 1.82)  -- ({(\ycut+0.67)/-1.66}, \ycut);
    
    \coordinate (I) at (-1, 1);
    
    \coordinate (A) at (1.5, 0.5);
    \coordinate (B) at (1.8, 0.8);
    \coordinate (C) at (2.1, 1.1);
    \coordinate (D) at (2.5, 1.5);
    
    \fill[C0] (A) circle (3pt);
    \fill[white] (A) circle (2.5pt);
    \fill[C0] (A) circle (1.5pt);
    
    \fill[C1] (B) circle (3pt);
    \fill[white] (B) circle (2.5pt);
    \fill[C1] (B) circle (1.5pt);
    
    \fill[C2] (C) circle (3pt);
    \fill[white] (C) circle (2.5pt);
    \fill[C2] (C) circle (1.5pt);
    
    \fill[C3] (D) circle (3pt);
    \fill[white] (D) circle (2.5pt);
    \fill[C3] (D) circle (1.5pt);

    \fill[black] (I) circle (3pt);
    \fill[white] (I) circle (2.5pt);
    \fill[black] (I) circle (1.5pt);

    \node[above right=4pt, font=\normalsize] at (I) {$\mathbf{L}^\star$};

    \node[C0, below right=2pt] at (A) {$\mathbf{A}$};
    \node[C1, below right=2pt] at (B) {$\mathbf{B}$};
    \node[C2, below right=2pt] at (C) {$\mathbf{C}$};
    \node[C3, below right=2pt] at (D) {$\mathbf{D}$};

    \node[C0, above left] at ({(\ycut+2)/-3}, \ycut) {$\mathbf{A}^\star$};
    \node[C1, below left] at ({(\ycut+1.25)/-2.25}, \ycut) {$\mathbf{B}^\star$};
    \node[C2, below right] at ({(\ycut+0.91)/-1.91}, \ycut) {$\mathbf{C}^\star$};
    \node[C3, above right] at ({(\ycut+0.67)/-1.66}, \ycut) {$\mathbf{D}^\star$};

\end{tikzpicture}
\caption{Secant configuration: $\mathbf{L}$ does not pass through the origin, dual hyperplanes meet at a finite point $\mathbf{L}^\star$.}
\label{fig:pga-duality-secant}
\end{subfigure}
 
\caption{Point-hyperplane duality in 2D PGA. Four collinear points on line $\mathbf{L}$ correspond to four dual hyperplanes (lines). The origin lies on $\mathbf{L}$ in (a) but not in (b), resulting in parallel versus secant dual hyperplanes. In both cases, the cross-ratio is preserved under duality.}
\label{fig:pga-duality}
\end{figure}

\begin{remark}
Under this correspondence, the unique ideal point allowed on the line maps exactly to the unique hyperplane passing through the origin, which forces the hyperplanes intersection to not pass through the origin. In the degenerate case where one of the four points is ideal, the corresponding dual hyperplane passes through the origin, and the cross-ratio degenerates to the affine ratio~\eqref{eq:pga_affine_ratio}, consistently with the classical theory of \cref{sec:proj-cr}.
\end{remark}
Each regressive product of points dualizes as
\begin{equation}
  \mathbf{P}_i \vee \mathbf{P}_j
  = \boldsymbol{\Pi}_i^\star \vee \boldsymbol{\Pi}_j^\star
  = \bigl(\boldsymbol{\Pi}_i \wedge \boldsymbol{\Pi}_j\bigr)^\star,
\end{equation}
which motivates the following definition, derived from \cref{eq:cr-points-pga} by duality.
\begin{definition}
\label{def:cr-finite-dual-hyperplanes-pga}
Let $\boldsymbol{\Pi}_1, \boldsymbol{\Pi}_2, \boldsymbol{\Pi}_3, \boldsymbol{\Pi}_4$ be four distinct hyperplanes in $n$-dimensional PGA sharing a common grade-$2$ flat intersection that does not pass through the origin. Their \emph{cross-ratio} is
\begin{equation}
  \{\boldsymbol{\Pi}_1, \boldsymbol{\Pi}_2;
    \boldsymbol{\Pi}_3, \boldsymbol{\Pi}_4\}
  =
  \frac{(\boldsymbol{\Pi}_1^\star \vee \boldsymbol{\Pi}_3^\star)
        (\boldsymbol{\Pi}_2^\star \vee \boldsymbol{\Pi}_4^\star)}
       {(\boldsymbol{\Pi}_1^\star \vee \boldsymbol{\Pi}_4^\star)
        (\boldsymbol{\Pi}_2^\star \vee \boldsymbol{\Pi}_3^\star)}.
  \label{eq:cr-dual-hyperplanes}
\end{equation}
\end{definition}
\begin{remark}
We adopt the regressive form rather than the wedge to preserve the duality correspondence with \eqref{eq:cr-points-pga}.
\end{remark}
Indeed, by construction, when $\boldsymbol{\Pi}_i = \mathbf{P}_i^\star$,
\eqref{eq:cr-points-pga} coincides with \eqref{eq:cr-dual-hyperplanes}:
\begin{equation}
  \{\boldsymbol{\Pi}_1, \boldsymbol{\Pi}_2;
    \boldsymbol{\Pi}_3, \boldsymbol{\Pi}_4\}
  = \{\mathbf{P}_1, \mathbf{P}_2; \mathbf{P}_3, \mathbf{P}_4\}.
\end{equation}

\subsection{Ideal Points Cross-Ratio}
\label{sec:ideal-cr}

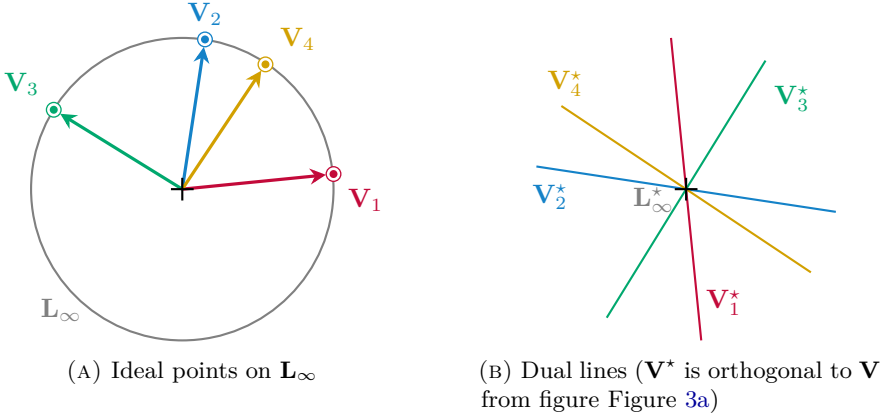
\begin{figure}[ht]
\centering
 
\begin{subfigure}[t]{0.45\textwidth}
\centering
\begin{tikzpicture}[scale=1]

    \coordinate (O) at (0,0);

    \draw[gray, thick] (0,0) circle (2);
    \node[gray] at (-1.6,-1.6) {$\mathbf{L}_\infty$};

    \coordinate (A) at (2,0.2);
    \coordinate (B) at (0.3,1.98);
    \coordinate (C) at (-1.7,1.05);
    \coordinate (D) at (1.1,1.65);

    \draw[C0, line width=1.2pt, -{Stealth[length=6pt, width=6pt]}] (O) -- ($(O)!0.95!(A)$);
    \draw[C1, line width=1.2pt, -{Stealth[length=6pt, width=6pt]}] (O) -- ($(O)!0.95!(B)$);
    \draw[C2, line width=1.2pt, -{Stealth[length=6pt, width=6pt]}] (O) -- ($(O)!0.95!(C)$);
    \draw[C3, line width=1.2pt, -{Stealth[length=6pt, width=6pt]}] (O) -- ($(O)!0.95!(D)$);

    \fill[C0] (A) circle (3pt);
    \fill[white] (A) circle (2.5pt);
    \fill[C0] (A) circle (1.5pt);

    \fill[C1] (B) circle (3pt);
    \fill[white] (B) circle (2.5pt);
    \fill[C1] (B) circle (1.5pt);

    \fill[C2] (C) circle (3pt);
    \fill[white] (C) circle (2.5pt);
    \fill[C2] (C) circle (1.5pt);

    \fill[C3] (D) circle (3pt);
    \fill[white] (D) circle (2.5pt);
    \fill[C3] (D) circle (1.5pt);

    \node[C0, below right=2pt] at (A) {$\mathbf{V}_1$};
    \node[C1, above =2pt] at (B) {$\mathbf{V}_2$};
    \node[C2, above left=2pt] at (C) {$\mathbf{V}_3$};
    \node[C3, above right=2pt] at (D) {$\mathbf{V}_4$};

    \draw[black, thick] (-0.15,0) -- (0.15,0);
    \draw[black, thick] (0,-0.15) -- (0,0.15);

\end{tikzpicture}
\caption{Ideal points on $\mathbf{L}_\infty$}
\label{fig:ideal-points}
\end{subfigure}
\hfill
\begin{subfigure}[t]{0.45\textwidth}
\centering
\begin{tikzpicture}[scale=1.0]

    \coordinate (O) at (0,0);

    \coordinate (A) at (2,0.2);
    \coordinate (B) at (0.3,1.98);
    \coordinate (C) at (-1.7,1.05);
    \coordinate (D) at (1.1,1.65);

    \draw[C0, thick] (-0.2,2) -- (0.2,-2);
    \node[C0] at (0.5,-1.5) {$\mathbf{V}_1^\star$};

    \draw[C1, thick] (-1.98,0.3) -- (1.98,-0.3);
    \node[C1] at (-1.8,-0.1) {$\mathbf{V}_2^\star$};

    \draw[C2, thick] (-1.05,-1.7) -- (1.05,1.7);
    \node[C2] at (1.4,1.2) {$\mathbf{V}_3^\star$};

    \draw[C3, thick] (-1.65,1.1) -- (1.65,-1.1);
    \node[C3] at (-1.6,1.4) {$\mathbf{V}_4^\star$};

    \draw[black, thick] (-0.15,0) -- (0.15,0);
    \draw[black, thick] (0,-0.15) -- (0,0.15);

    \node[gray, below=4pt, left=1pt] at (0,0) {$\mathbf{L}_\infty^\star$};

\end{tikzpicture}
\caption{Dual lines ($\mathbf{V}^\star$ is orthogonal to $\mathbf{V}$  from figure \cref{fig:ideal-points})}
\label{fig:dual-lines}
\end{subfigure}
 
\caption{Ideal points duality in 2D PGA. Ideal directions correspond to points on $\mathbf{L}_\infty$, while their duals are orthogonal lines through the origin. More generally, it extends to hyperplanes in $n$-d.}
\label{fig:pga_duality_ideal_points}
\end{figure}

A hyperplane passing through the origin has vanishing ideal part and is therefore the Hodge dual of an ideal point. This is precisely the configuration excluded by the hypothesis chosen in~\eqref{eq:cr-dual-hyperplanes}. The present section defines the cross-ratio for four such ideal points aligned on an ideal line $\mathbf{L}_\infty$, playing the same role as \cref{sec:classicalcrossratio} did for finite points, and derives the corresponding origin-passing hyperplane cross-ratio by duality. \cref{fig:pga_duality_ideal_points} illustrates this duality link.

\begin{definition}
Let $\mathbf{V}_1, \mathbf{V}_2, \mathbf{V}_3, \mathbf{V}_4$ be four distinct ideal points on the same ideal line $\mathbf{L}_\infty$. Their \emph{ideal cross-ratio} is
\begin{equation}
  \{\mathbf{V}_1, \mathbf{V}_2; \mathbf{V}_3, \mathbf{V}_4\}
  = \frac{(\mathbf{V}_1^\star \wedge \mathbf{V}_3^\star)
          (\mathbf{V}_2^\star \wedge \mathbf{V}_4^\star)}
         {(\mathbf{V}_1^\star \wedge \mathbf{V}_4^\star)
          (\mathbf{V}_2^\star \wedge \mathbf{V}_3^\star)}.
  \label{eq:cr-ideal-points-pga}
\end{equation}
\end{definition}
Each dual $\mathbf{V}_i^\star = \boldsymbol{\Pi}_i$ is a grade-$1$ Euclidean vector, i.e.\ a hyperplane through the origin. Their wedge product is proportional to their intersection, a Euclidean grade-$2$ object encoding the dual of the ideal support line $\mathbf{L}_\infty$, with scalar factor given by the sine of the angle:
\begin{equation}
  \mathbf{V}_i^\star \wedge \mathbf{V}_j^\star
  = \boldsymbol{\Pi}_i \wedge \boldsymbol{\Pi}_j
  = \|\mathbf{V}_i\|\,\|\mathbf{V}_j\|\,\sin(\alpha_{ij})\,\hat{\mathbf{L}}_\infty^\star,
  \qquad \alpha_{ij} = \alpha_i - \alpha_j,
\end{equation}
where $(\hat{\mathbf{L}}_\infty^\star)^2 = \pm \mathbf{1}$, which cancels in the
ratio~\eqref{eq:cr-ideal-points-pga}, recovering the sine cross-ratio
of \cref{sec:sine-cr}:
\begin{equation}
  \{\mathbf{V}_1, \mathbf{V}_2; \mathbf{V}_3, \mathbf{V}_4\}
  = \frac{\sin(\alpha_{13})\,\sin(\alpha_{24})}
         {\sin(\alpha_{14})\,\sin(\alpha_{23})}.
\end{equation}

The origin-passing hyperplane cross-ratio follows by duality. Setting $\boldsymbol{\Pi}_i = \mathbf{V}_i^\star$, the wedge $\boldsymbol{\Pi}_i \wedge \boldsymbol{\Pi}_j$ is a pure Euclidean blade, so no additional dualization is needed.

\begin{definition}
\label{def:cr-origin-hyperplanes-pga}
Let $\boldsymbol{\Pi}_1, \boldsymbol{\Pi}_2, \boldsymbol{\Pi}_3, \boldsymbol{\Pi}_4$ be four distinct hyperplanes sharing a common grade-$2$ intersection that passes through the origin. Their \emph{cross-ratio} is
\begin{equation}
  \{\boldsymbol{\Pi}_1, \boldsymbol{\Pi}_2;
    \boldsymbol{\Pi}_3, \boldsymbol{\Pi}_4\}
  =
  \frac{(\boldsymbol{\Pi}_1 \wedge \boldsymbol{\Pi}_3)
        (\boldsymbol{\Pi}_2 \wedge \boldsymbol{\Pi}_4)}
       {(\boldsymbol{\Pi}_1 \wedge \boldsymbol{\Pi}_4)
        (\boldsymbol{\Pi}_2 \wedge \boldsymbol{\Pi}_3)}.
  \label{eq:cr-ideal-dual-hyperplanes}
\end{equation}
\end{definition}
By construction, when $\boldsymbol{\Pi}_i = \mathbf{V}_i^\star$, \eqref{eq:cr-ideal-points-pga} coincides with \eqref{eq:cr-ideal-dual-hyperplanes}, so the cross-ratio is preserved under the ideal point hyperplane duality.

\subsection{Finite Flats Cross-Ratio}
\label{sec:finite_flat_cross_ratio}
For intermediate grade-$k$ blades with $1 < k < n$, the wedge and the regressive products vanish for two co-planar/secant objects. In this case, the commutator need to be used to extracts the natural scalar measurement. In the case of object sharing a common intersection the commutator extract a bivector blade that encode the pencil generator of the pairs of object, this is what we used in the folowing.

\begin{definition}
\label{def:cr-parallel-flats}
Let $\mathbf{F}_1, \mathbf{F}_2, \mathbf{F}_3, \mathbf{F}_4$ be four distinct grade-$k$ flats in $n$-dimensional PGA sharing a common ideal intersection $A_\infty$. Their \emph{cross-ratio} is
\begin{equation}
  \{\mathbf{F}_1, \mathbf{F}_2; \mathbf{F}_3, \mathbf{F}_4\}
  =
  \frac{(\mathbf{F}_1 \times^\star \mathbf{F}_3)
        (\mathbf{F}_2 \times^\star \mathbf{F}_4)}
       {(\mathbf{F}_1 \times^\star \mathbf{F}_4)
        (\mathbf{F}_2 \times^\star \mathbf{F}_3)}.
  \label{eq:cr-parallel-flats}
\end{equation}
\end{definition}
Since each $\mathbf{F}_i$ and $\mathbf{F}_j$ are parallel, their commutator $\mathbf{F}_i \times \mathbf{F}_j$ is a grade-$2$ ideal blade proportional to the unit ideal bivector $\hat{\mathbf{B}}_\infty$, with scalar factor equal to the signed parameter difference $t_i-t_j$ along the hypervolume where they lie
\begin{equation}
  \mathbf{F}_i \times \mathbf{F}_j
  = \| \mathbf{F}_i\| \| \mathbf{F}_j\| (t_i-t_j) \hat{\mathbf{B}}_\infty
\end{equation}
where $(\hat{\mathbf{B}}_\infty)^2=0$ because it is a pure ideal bivector. We use the dual to pass to a non ideal element

\begin{equation}
  (\mathbf{F}_i \times \mathbf{F}_j)^\star
  = \| \mathbf{F}_i\| \| \mathbf{F}_j\| (t_i-t_j) \hat{\mathbf{B}}_\infty^\star
\end{equation}
where $(\hat{\mathbf{B}}_\infty^\star)^2= \pm 1$. To avoid the sourouding dual and to use a single operator we rewrite it as the commutator dual
\begin{equation}
  (\mathbf{F}_i \times \mathbf{F}_j)^\star
  = \mathbf{F}_i^\star \times^\star \mathbf{F}_j^\star
\end{equation}

\begin{remark}
The condition that at most one flat passes through the origin mirrors the condition in \eqref{eq:cr-dual-hyperplanes}: a flat through the origin has no $\mathbf{e}_0$ component in its Euclidean part, and its Hodge dual is an ideal flat, which is the configuration treated in \cref{sec:ideal_flat_cross_ratio}.
\end{remark}

The Hodge dual maps each finite grade-$k$ flat $\mathbf{F}_i$ to a grade-$(n+1-k)$ flat~$\mathbf{F}_i^\star$. When $\mathbf{F}_i$ and $\mathbf{F}_j$ are parallel finite flats sharing a common ideal intersection, their duals $\mathbf{F}_i^\star$ and $\mathbf{F}_j^\star$ are secant finite flats sharing a common finite intersection not through the origin. This is the exact dual counterpart of the parallel configuration, just as duality exchanges finite points and parallel hyperplanes in \cref{sec:dual-hyperplane}. Using the identity $A \times^\star B = (A^\star \times B^\star)^\star$, this motivates the following definition, derived from \eqref{eq:cr-parallel-flats} by duality.

\begin{definition}
\label{def:cr-finite-flats}
Let $\mathbf{F}_1, \mathbf{F}_2, \mathbf{F}_3, \mathbf{F}_4$ be four distinct finite grade-$k$ flats in $n$-dimensional PGA sharing a common finite intersection, at most one of which passes through the origin. Their \emph{cross-ratio} is
\begin{equation}
  \{\mathbf{F}_1, \mathbf{F}_2; \mathbf{F}_3, \mathbf{F}_4\}
  =
  \frac{(\mathbf{F}_1^\star \times^\star \mathbf{F}_3^\star)
        (\mathbf{F}_2^\star \times^\star \mathbf{F}_4^\star)}
       {(\mathbf{F}_1^\star \times^\star \mathbf{F}_4^\star)
        (\mathbf{F}_2^\star \times^\star \mathbf{F}_3^\star)}.
  \label{eq:cr-finite-flats}
\end{equation}
\end{definition}

Since $\mathbf{F}_i$ and $\mathbf{F}_j$ are secant and finite and do not both pass through the origin, their commutator is proportional to a fixed blade $\mathbf{B}$ encoding the associated rotation axis:
\begin{equation}
  \mathbf{F}_i \times \mathbf{F}_j
  = \pm\sin(\alpha_{ij})\,\mathbf{B},
  \qquad \alpha_{ij} = \alpha_i - \alpha_j.
\end{equation}
The blade $\mathbf{B}$ has both a Euclidean part $\mathbf{B}_E$ and an ideal part $\mathbf{e}_0\mathbf{B}_I$, encoding respectively the direction and the moment of the rotation axis. Because of this mixed structure, $\mathbf{B}^2 \neq 0$: the cross terms between the Euclidean and ideal parts contribute a non-vanishing scalar. Its dual $\mathbf{B}^\star$ shares the same property and cancels as a common factor in all four terms of~\eqref{eq:cr-finite-flats}, yielding the sine cross-ratio of \cref{sec:sine-cr}.

By construction, when $\mathbf{F}_i = \mathbf{G}_i^\star$ for parallel finite flats $\mathbf{G}_i$, \eqref{eq:cr-finite-flats} coincides with \eqref{eq:cr-parallel-flats}, so the cross-ratio is preserved under the parallel--secant duality for intermediate flats.

\subsection{Ideal Flats Cross-Ratio}
\label{sec:ideal_flat_cross_ratio}

The Hodge dual of an ideal grade-$k$ flat is a finite grade-$(n+1-k)$ flat passing through the origin. This is precisely the configuration excluded by the hypothesis chosen in \eqref{eq:cr-finite-flats} and \eqref{eq:cr-parallel-flats}. The present section defines the cross-ratio for ideal flats, playing the same role as \cref{sec:ideal-cr} did for ideal points.

\begin{definition}
\label{def:cr-ideal-flats}
Let $\mathbf{F}_1, \mathbf{F}_2, \mathbf{F}_3, \mathbf{F}_4$ be four distinct ideal grade-$k$ flats in $n$-dimensional PGA sharing a common ideal intersection. Their \emph{cross-ratio} is
\begin{equation}
  \{\mathbf{F}_1, \mathbf{F}_2; \mathbf{F}_3, \mathbf{F}_4\}
  =
  \frac{(\mathbf{F}_1^\star \times \mathbf{F}_3^\star)
        (\mathbf{F}_2^\star \times \mathbf{F}_4^\star)}
       {(\mathbf{F}_1^\star \times \mathbf{F}_4^\star)
        (\mathbf{F}_2^\star \times \mathbf{F}_3^\star)}.
  \label{eq:cr-ideal-flats}
\end{equation}
\end{definition}

Because $\mathbf{F}_i$ and $\mathbf{F}_j$ are both ideal, their geometric product vanishes ($\mathbf{e}_0^2 = 0$), and so does their commutator directly. The signed measurement is instead extracted by dualizing the operands first: the duals $\mathbf{F}_i^\star$ are finite grade-$(n+1-k)$ flats passing through the origin, and their commutator is proportional to a fixed Euclidean blade $\mathbf{B}_E$ encoding the associated rotation axis:
\begin{equation}
  \mathbf{F}_i^\star \times \mathbf{F}_j^\star
  = \pm\sin(\alpha_{ij})\,\mathbf{B}_E,
  \qquad \alpha_{ij} = \alpha_i - \alpha_j,
\end{equation}
with $\mathbf{B}_E^2 = \pm\mathbf{1}$, which cancels in the ratio~\eqref{eq:cr-ideal-flats}. This definition is the direct counterpart of \eqref{eq:cr-finite-flats} applied to the dual objects, and mirrors the ideal point cross-ratio \eqref{eq:cr-ideal-points-pga} for intermediate grades.

By the same duality argument as above, the Hodge dual of two secant ideal flats are origin-passing finite flats whose commutator dual encodes their signed distance. This motivates the following definition, derived from \eqref{eq:cr-ideal-flats} by duality.

\begin{definition}
\label{def:cr-origin-flats}
Let $\mathbf{F}_1, \mathbf{F}_2, \mathbf{F}_3, \mathbf{F}_4$ be four distinct finite grade-$k$ flats in $n$-dimensional PGA all passing through the origin. Their \emph{cross-ratio} is
\begin{equation}
  \{\mathbf{F}_1, \mathbf{F}_2; \mathbf{F}_3, \mathbf{F}_4\}
  =
  \frac{(\mathbf{F}_1 \times \mathbf{F}_3)
        (\mathbf{F}_2 \times \mathbf{F}_4)}
       {(\mathbf{F}_1 \times \mathbf{F}_4)
        (\mathbf{F}_2 \times \mathbf{F}_3)}.
  \label{eq:cr-origin-flats}
\end{equation}
\end{definition}

Because the flats pass through the origin they have no $\mathbf{e}_0$ component in their Euclidean part, so their commutator is a purely Euclidean blade $\mathbf{B}_E$ with $\mathbf{B}_E^2 = \pm\mathbf{1}$, which cancels in the ratio~\eqref{eq:cr-origin-flats}. By construction, when $\mathbf{F}_i = \mathbf{G}_i^\star$ for ideal flats $\mathbf{G}_i$, \eqref{eq:cr-origin-flats} coincides with \eqref{eq:cr-ideal-flats}, so the cross-ratio is preserved under the ideal--origin-passing duality for intermediate flats.

\subsection{Recap}
\label{sec:recap}
The previous subsections have established cross-ratio formulas for all object types in $n$-dimensional PGA: finite and ideal points (\cref{sec:classicalcrossratio,sec:ideal-cr}), hyperplanes (\cref{sec:dual-hyperplane}), and intermediate flats (\cref{sec:finite_flat_cross_ratio,sec:ideal_flat_cross_ratio}). Although all formulas yield the correct projective invariant, they are not expressed in terms of the same operator. \Cref{tab:cr-formulas} collects the canonical pairwise measurement operator for each configuration.

\begin{table}[H]
\centering
\caption{Pairwise measurement operator for each object type and configuration in $n$-dimensional PGA. Lines with identical colors show dual configurations.}
\label{tab:cr-formulas}
\setlength{\tabcolsep}{5pt}
\begin{tabular}{@{}llll@{}}
\toprule
Object & Grade & Support/Configuration & Operator \\
\midrule
\rowcolor{C0!15}
Finite points  & $n$ & finite support $\mathbf{L}$
  & $\mathbf{P}_i \vee \mathbf{P}_j$ \\
\rowcolor{C1!15}
Ideal points   & $n$ & ideal support $\mathbf{L}_\infty$
  & $\mathbf{V}_i^\star \wedge \mathbf{V}_j^\star$ \\
\midrule
\rowcolor{C0!15}
Hyperplanes    & $1$ & intersection not through origin
  & $\boldsymbol{\Pi}_i^\star \vee \boldsymbol{\Pi}_j^\star$ \\
\rowcolor{C1!15}
Hyperplanes    & $1$ & intersection through origin
  & $\boldsymbol{\Pi}_i \wedge \boldsymbol{\Pi}_j$ \\
\midrule
\rowcolor{C2!15}
Flats          & $k$ & finite intersection not through origin
  & $\mathbf{F}_i^\star \times^\star \mathbf{F}_j^\star$ \\
\rowcolor{C3!15}
Flats          & $k$ & intersection through origin
  & $\mathbf{F}_i \times \mathbf{F}_j$ \\
\rowcolor{C2!15}
Finite flats   & $k$ & ideal intersection (parallel)
  & $\mathbf{F}_i \times^\star \mathbf{F}_j$ \\
\rowcolor{C3!15}
Ideal flats    & $k$ & ideal intersection (secant)
  & $\mathbf{F}_i^\star \times \mathbf{F}_j^\star$ \\
\bottomrule
\end{tabular}
\end{table}

The four observations from \cref{tab:cr-formulas} reveal a clear
duality structure.
Despite this elegant duality structure, no single operator from \cref{tab:cr-formulas} works uniformly across all eight configurations. Moreover, even within the same grade, different configurations require different operators. This fragmentation stems from the degenerate metric of PGA: the null vector $\mathbf{e}_0$ breaks the symmetry between finite and ideal elements, so that operators which work on finite objects fail on ideal ones, and the origin plays a distinguished role that has no classical counterpart.

To unify the signed measurement operator across all configurations, we first observe that for grade-$1$ objects the wedge and commutator
coincide:
\begin{equation}
  \boldsymbol{\Pi}_i \wedge \boldsymbol{\Pi}_j
  = \boldsymbol{\Pi}_i \times \boldsymbol{\Pi}_j,
\end{equation}
since for two 1-vectors the antisymmetric part of the geometric product is precisely the wedge. Similarly, for grade-$n$ objects, the regressive product and the commutator dual coincide:
\begin{equation}
  \mathbf{P}_i \vee \mathbf{P}_j
  = \mathbf{P}_i \times^\star \mathbf{P}_j,
\end{equation}
since the regressive product is the dualized wedge, and the commutator dual is the dualized commutator, and at grade $n$ the two coincide. Substituting these identities into \cref{tab:cr-formulas}, all eight operators reduce to either $\times$ or $\times^\star$ :

\begin{table}[H]
\centering
\caption{Unified pairwise measurement operators after substituting
the wedge--commutator and regressive--commutator dual identities.}
\label{tab:cr-formulas-unified}
\setlength{\tabcolsep}{5pt}
\begin{tabular}{@{}llll@{}}
\toprule
Object & Grade & Support/Configuration & Operator \\
\midrule
\rowcolor{C0!15}
Finite points  & $n$ & finite support $\mathbf{L}$
  & $\mathbf{P}_i \times^\star \mathbf{P}_j$ \\
\rowcolor{C1!15}
Ideal points   & $n$ & ideal support $\mathbf{L}_\infty$
  & $\mathbf{V}_i^\star \times \mathbf{V}_j^\star$ \\
\midrule
\rowcolor{C0!15}
Hyperplanes    & $1$ & intersection not through origin
  & $\boldsymbol{\Pi}_i^\star \times^\star \boldsymbol{\Pi}_j^\star$ \\
\rowcolor{C1!15}
Hyperplanes    & $1$ & intersection through origin
  & $\boldsymbol{\Pi}_i \times \boldsymbol{\Pi}_j$ \\
\midrule
\rowcolor{C2!15}
Flats          & $k$ & finite intersection not through origin
  & $\mathbf{F}_i^\star \times^\star \mathbf{F}_j^\star$ \\
\rowcolor{C3!15}
Flats          & $k$ & intersection through origin
  & $\mathbf{F}_i \times \mathbf{F}_j$ \\
\rowcolor{C2!15}
Finite flats   & $k$ & ideal intersection (parallel)
  & $\mathbf{F}_i \times^\star \mathbf{F}_j$ \\
\rowcolor{C3!15}
Ideal flats    & $k$ & ideal intersection (secant)
  & $\mathbf{F}_i^\star \times \mathbf{F}_j^\star$ \\
\bottomrule
\end{tabular}
\end{table}

Despite this unified view, two distinct operators $\times$ and
$\times^\star$ are still needed, and the choice between them depends
on whether the objects are finite or ideal and whether their common
intersection passes through the origin. The remaining fragmentation
still reflects the degenerate metric of PGA.

\section{Conclusion}
\label{sec:conclusion}

This paper has developed a complete theory of projective cross-ratios
in $n$-dimensional Plane-Based Geometric Algebra $\mathbb{R}_{n,0,1}$,
covering geometric objects of every grade. For each object type and
configuration we derived an explicit cross-ratio formula and
identified the canonical pairwise measurement operator. In every case,
the formula recovers the appropriate classical invariant: signed
distance ratios for parallel configurations and sine cross-ratios for
secant ones.

A systematic duality analysis revealed that all eight configurations
organize into four dual pairs under the Hodge dual.
All measurement operators across
the eight configurations reduce to either the commutator $\times$ or
the commutator dual $\times^\star$. The remaining dependence on two
operators rather than one reflects an intrinsic feature of the
degenerate PGA metric: the null vector $\mathbf{e}_0$ breaks the
symmetry between finite and ideal elements in a way that no single
standard operator can eliminate.

These results establish the cross-ratio as a grade-agnostic projective
invariant within PGA and provide a constructive algebraic foundation
for defining $n$-dimensional homographies from prescribed invariants.
Future work will address the further unification of these operators,
the inverse problem of reconstructing geometric objects from a given
cross-ratio value, and the application of these invariants to
construct homographies within the PGA framework.

\bibliographystyle{spmpsci}
\bibliography{references}

\end{document}